\documentclass[preprintnumbers, floatfix, showkeys, preprintnumbers, letterpaper, twocolumn, superscriptaddress,nofootinbib]{revtex4-2}
\pdfoutput=1 
\usepackage{graphicx}
\usepackage{microtype}
\usepackage{amsmath}
\usepackage{amssymb}
\usepackage{subfigure}
\usepackage{url}
\usepackage{hyperref}
\usepackage{mathtools}
\usepackage{orcidlink}
\IfFileExists{slashbox.sty}{\usepackage{slashbox}}{} 

\usepackage{xcolor}
\usepackage{color}
\usepackage{mathrsfs}
\usepackage{calrsfs}
\usepackage{amsfonts}
\usepackage{tabularx}
\usepackage{latexsym}
\usepackage{ragged2e}
\usepackage{epsfig}
\usepackage{textcomp}
\usepackage{float}

\usepackage{caption}
\DeclareCaptionJustification{justified}{\leftskip=0pt \rightskip=0pt \parfillskip=0pt plus 1fil}
\definecolor{vividviolet}{rgb}{0.62, 0.0, 1.0}
\definecolor{amaranth}{rgb}{0.9, 0.17, 0.31}
\definecolor{palatinateblue}{rgb}{0.15, 0.23, 0.89}
\definecolor{brightpink}{rgb}{1.0, 0.0, 0.5}
\definecolor{cornflowerblue}{rgb}{0.39, 0.58, 0.93}
\definecolor{deepcarminepink}{rgb}{0.94, 0.19, 0.22}
\definecolor{radicalred}{rgb}{1.0, 0.21, 0.37}

\hypersetup{ linktoc=all,
	colorlinks, linkcolor={palatinateblue},
	citecolor={brightpink}, urlcolor={amaranth}
}

\graphicspath{{Images/}}

\def\sideremark#1{\ifvmode\leavevmode\fi\vadjust{\vbox to0pt{\vss
			\hbox to 0pt{\hskip\hsize\hskip1em
				\vbox{\hsize1.3cm\tiny\raggedright\pretolerance10000
					\noindent #1\hfill}\hss}\vbox to8pt{\vfil}\vss}}}%
\def\beq{\begin{equation}}
\def\eeq{\end{equation}}

\newcommand{\BH}{S_{\mathrm{BH}}}

\newcommand{\ee}{\mathrm{e}}
\newcommand{\mi}{I}  

\begin{document}
\title{Daemon sub Ponte: The Cost of a Second Black Hole \\(A Scholium on an ER=EPR Thought Experiment)}

\author{Yen Chin \surname{Ong}\orcidlink{0000-0002-3944-1693}}
\email{ongyenchin@nuaa.edu.cn}
\affiliation{Center for the Cross-disciplinary Research of Space Science and Quantum-technologies (CROSS-Q), College of Physics, Nanjing University of Aeronautics and Astronautics, \\29 Jiangjun Road, Nanjing City, Jiangsu Province 211106, China}

\begin{abstract}
A thought experiment in the ER=EPR and related literature involves collecting the Hawking radiation of an old black hole, then collapsing it into a second, entangled black hole. 
In this short note, we examine the resources required by this idealized conversion. More specifically, we are interested in answering the question: how much external energy is needed to form the second black hole?  
Assuming a semiclassical bound on the information capacity of the newly formed black hole, we derive a conditional lower bound on external energy and a complementary constraint on correlations retained by the exterior degrees of freedom. We also distinguish entropy-level feasibility from the requirements for exact and approximate purification. We illustrate these requirements via a Page curve analysis.
\end{abstract}

\maketitle

\section{Introduction}
Maldacena and Susskind proposed that the entanglement of two black holes may admit an Einstein--Rosen bridge interpretation ~\cite{MaldacenaSusskind}. This is known as the ``ER=EPR'' conjecture. 
In that paper, they also proposed a thought experiment in which an observer (Alice) collects the Hawking radiation and converts it into a second black hole with an exceptionally powerful quantum computer. For an evaporating black hole in asymptotically flat spacetime, the observer may then act on the second black hole to prepare a thermofield-double state, subject to suitable assumptions about the reduced density matrix of the first black hole. Note that this is an idealization of quantum control, not a detailed dynamical or thermodynamic construction. In fact, this thought experiment serves primarily as a \emph{conceptual illustration} of ER=EPR, and is not meant as a detailed proposal for how to physically construct an Einstein--Rosen bridge (hereinafter, ``ER-bridge''). Still, it would be interesting to study more carefully what the thought experiment entails.

The conversion of collected Hawking radiation into a second black hole has continued to appear in the literature, especially in discussions of information recovery and the computational complexity of decoding \cite{MaldacenaStanfordYang,Boeretal,Brownetal}. In particular, Maldacena, Stanford and Yang \cite{MaldacenaStanfordYang} employed it in a geometric interpretation of information recovery. It is worth noting that Brown et al. explicitly emphasized that the thought experiment is illustrative rather than practical \cite{Brownetal}. Here we examine a complementary question: the energy and purification requirements of the conversion itself, independently of the computational complexity of implementing it.

Our question is whether the available Hawking radiation can form a second black hole large enough to accommodate the purification of the first. Differences between coarse-grained horizon entropies alone do not determine the entropy of an ER bridge. The purpose of this note is to identify precisely what is required if this illustration is recast as a physical quantum preparation protocol. It is important to distinguish among (i) the formation of a second black hole, (ii) transfer of the \emph{full purification} of the surviving black hole into that second black hole, and (iii) preparation of a special two-sided state capable of admitting a smooth semiclassical bridge. Only the second task is directly constrained by the elementary entropy capacity considered below. More information about the state is required if we want to determine (iii). 

Indeed, the physical realizability of black hole information thought experiments need not follow from their formal consistency. For example, Harlow and Hayden argued that even with an extremely powerful quantum computer, decoding Hawking radiation may require computational timescales far exceeding the lifetime of the black hole \cite{HarlowHayden}. In this work, we shall examine a complementary constraint: the energy and entropy bookkeeping involved in assembling
the collected radiation into a second black hole. Our investigation is also complementary to the analysis of entanglement conservation and wormhole geometry~\cite{RemmenBaoPollack}.

We assume unitary evolution of the complete system and adopt the standard semiclassical Bekenstein-Hawking entropy relation \cite{Bekenstein}, according to which the logarithm of the number of black-hole microstates (equivalently, the logarithm of the dimension of the effective Hilbert space) is proportional to the area of the event horizon. We treat this relation as an approximate entropy-capacity bound. The information-theoretic inequalities used here are standard; we are not proposing a new entropy bound. Our objective is to derive explicitly the consequences of these assumptions for the proposed operation, including the often overlooked roles of external energy and residual purifying degrees of freedom.

\section{A conditional quantum preparation protocol}
We shall work in four-dimensional asymptotically flat spacetime\footnote{We use the natural units in which $G=\hbar=c=k_{\mathrm B}=1$.}. Consider an initially pure Schwarzschild black hole of mass $M_0$ that has evaporated to mass $M$. Neglecting energy losses during collection, the collected outgoing Hawking radiation $R$ carries energy
\begin{equation}
E_R\simeq M_0-M.
\label{ER}
\end{equation}
At a given time, let $A$ denote the remaining black hole and assume that the combined $AR$ state is pure, with Schmidt decomposition given by
\begin{equation}
|\Psi\rangle_{AR}=\sum_{i=1}^{r}\sqrt{p_i}\,|i\rangle_A|i\rangle_R, \quad p_1\geqslant p_2\geqslant\cdots>0.\label{Schmidt}
\end{equation}
Denote the (von Neumann) entanglement entropy by
\begin{equation}
s\equiv -\sum_{i=1}^r p_i \ln p_i =S(A)=S(R).
\end{equation}
The equality between $S(A)$ and $S(R)$ follows from the fact that reduced density matrices of a bipartite pure state have identical nonzero eigenvalues\footnote{This equality depends on the purity assumption. If $AR$ is mixed because some radiation or other purifying degrees of freedom have been excluded, then $S(A) \neq S(R)$ in general.}.
For simplicity, we shall neglect uncollected radiation and any independently entangled external reference; including them would require an enlarged system.

We model Alice's conversion of the collected radiation into a second black hole as an isometric transformation between the relevant Hilbert spaces\footnote{Here, isometry refers to preservation of the Hilbert space inner product, $V^\dagger V=\mathbb I_R$. The transformation can be viewed as part of a unitary evolution on an enlarged Hilbert space (a unitary operator is a surjective isometry).},
\begin{equation}
V:\mathcal H_R\longrightarrow\mathcal H_B\otimes\mathcal H_D.
\label{channel}
\end{equation}
Here $B$ designates the new black hole subsystem, and $D$ includes discarded modes, laboratory records, apparatus, any uncollapsed radiation, and all their purifiers. The overall $ABD$ state is assumed pure, and $A$ is untouched during Alice's operation. The \emph{clean} protocol demands that $D$ be decoupled and that $AB$ be pure, though a less demanding protocol can permit the original purification to be shared by $B$ and $D$. We emphasize that this transformation is an effective description of the quantum information transfer. It does \emph{not} describe the gravitational collapse dynamics. The distinction is important because an initially pure apparatus can also carry energy, so purity does not imply that Alice supplies no external energy. That contribution is accounted for separately through the quantity we call $\mathcal{E}_{\rm ext}$ later on.

We use two distinct inputs: semiclassical information capacity and energy accounting. First, in a suitable microcanonical energy window, a new Schwarzschild black hole of mass $m$ is assigned an effective information capacity\footnote{The microcanonical restriction refers to black-hole states within a narrow mass interval around $m$, with other macroscopic charges fixed, i.e., $d_B(m)$ denotes the effective number of microstates available within said interval. The microcanonical description is appropriate because we want to estimate the information capacity of a black hole with a specified mass, instead of its thermal properties in contact with a heat bath.}
\begin{equation}
\log d_B(m)\lesssim \BH(m)=4\pi m^2.
\label{capacity}
\end{equation}
This is an \emph{assumption} about the semiclassical microstate capacity, and does not consider the full gravitational Hilbert space. All resulting mass bounds are understood at leading semiclassical order.

This entropy-capacity assumption is not an application of the generalized second law (GSL). A GSL analysis would require specifying the relevant horizons and exterior entropies, but the radiation entanglement entropy $s$ cannot automatically be identified with the exterior entropy in that law~\cite{Wall}.

Secondly, let $\mathcal{E}_{\rm ext}\geqslant0$ denote the \emph{net additional energy made available to form} the second black hole $B$, 
beyond the energy $E_R$ contained in the collected Hawking radiation.
Energy accounting requires
\begin{equation}
m\leqslant E_R+\mathcal{E}_{\rm ext}.
\label{energy}
\end{equation}
This is an optimistic upper bound, since energy retained in other degrees of freedom can only reduce the final black hole mass. 

\section{Purification capacity and residual correlations}

If $AB$ is pure, $B$ must carry the entire purification of $A$.
Since operations on $R$ leave the reduced density matrix $\rho_A$
unchanged, all $r$ nonzero Schmidt coefficients in
Eq.\eqref{Schmidt} must be accommodated by $B$. Thus,
\begin{equation}
r\leqslant d_B(m).
\label{rank}
\end{equation}
This requirement is stronger than the often-quoted entropy
condition $s\leqslant \log d_B(m)$. Indeed, the von Neumann entropy
satisfies
\begin{equation}
s=-\sum_{i=1}^{r}p_i\log p_i\leqslant \log r,
\end{equation}
with equality only when all nonzero Schmidt probabilities
are equal. Consequently, $r\leqslant  d_B(m)$ implies
$s\leqslant \log d_B(m)$, but the converse need not hold.

To see this, consider a state with three nonzero Schmidt
probabilities,
\begin{equation}
p_1=1-2\delta,\qquad p_2=p_3=\delta,
\end{equation}
where $0<\delta \leqslant 1/3$. As $\delta\to0$, the entanglement
entropy approaches zero, yet the Schmidt rank\footnote{The underlying distinction is really this: the von Neumann entropy is only weakly affected by very small eigenvalues, whereas the Schmidt rank counts every strictly nonzero eigenvalue regardless of how tiny its probability is.} remains $r=3$. Since $r> 1$, the state remains entangled.
Thus, even when $s<\log 2$, a two-dimensional system $B$ cannot purify $A$.
This illustrates why an entropy bound alone cannot characterize exact purification, and motivates the finite-fidelity formulation developed below.

Without demanding a clean final state, purity of $ABD$ implies, in terms of mutual information,
\begin{equation}
\mi(A\!:\!B)+\mi(A\!:\!D)=2S(A)=2s,
\label{MIidentity}
\end{equation}
where $\mi(X\!:\!Y)=S(X)+S(Y)-S(XY)$. By the standard triangle inequality for von Neumann entropy (the Araki-Lieb inequality), we have $\mi(A\!:\!B)\leqslant2S(B)\leqslant2\log d_B(m)$. It follows that
\begin{equation}
\mi(A\!:\!D)\geqslant2\bigl[s-\BH(m)\bigr]_+, 
\quad [z]_+\equiv\max\{z,0\}.
\label{leakage}
\end{equation}
This tracks the \emph{correlations with the original black hole}, not merely the entropy present in the discarded apparatus. It does not assert that $D$ contains this fraction of the energy or of the emitted quanta. 
If $s>\BH(m)$, the second black hole has insufficient entropy capacity to accommodate the entanglement entropy of $A$,so $\mi(A\!:\!D)>0$. In this case \emph{some} correlations with $A$ must remain in $D$. When $s\leqslant \BH(m)$, however, $I(A:D) \geqslant 0$ does not impose any new constraint (mutual information is already nonnegative by subadditivity). In this case $D$ is not required to retain any correlations with $A$.

For a clean conversion, $AB$ is pure and $D$ decoupled, but $\mi(A\!:\!D)=0$ alone does not guarantee that $AB$ is pure. We note that
\begin{equation}
s\leqslant\BH(m)=4\pi m^2 \Longrightarrow m\geqslant \sqrt{\frac{s}{4\pi}}.
\label{clean}
\end{equation}
This entropy-level condition is necessary, but weaker than the exact rank requirement in Eq.\eqref{rank}.

Combining Eqs.\eqref{energy} and~\eqref{clean} gives the necessary ``energy-budget inequality''
\begin{equation}
\mathcal{E}_{\rm ext}\geqslant \mathcal{E}^{(S)}_{\rm{req}}\equiv\left[\sqrt{\frac{s}{4\pi}}-E_R\right]_+.
\label{energybound}
\end{equation}
The actual minimum energy for an exact implementation may exceed $\mathcal{E}^{(S)}_{\rm{req}}$ because gravitational dynamics and state preparation may impose further restrictions. Thus $\mathcal{E}^{(S)}_{\rm{req}}$ is an \emph{entropy-level lower bound}. The superscript $S$ emphasizes this entropy aspect. We should not confuse $\mathcal{E}^{(S)}_{\rm{req}}$ with the actual minimum energy required by a realizable formation protocol.

The exact Schmidt-rank requirement can be relaxed if one permits an \emph{approximate} rather than perfect purification. 
This would be arguably more realistic, even for our already rather idealized model.
We now quantify this relaxation in terms of the Schmidt spectrum.
Let $P_d=\sum_{i=1}^{d}p_i$ (with $P_d=1$ for $d\geqslant r$). Let $\rho_{AB}=\operatorname{Tr}_D
|\Phi\rangle_{ABD}\langle\Phi|$ denote the reduced state of the two black holes following Alice's operation on $R$, and let $|\phi\rangle_{AB}$ be any pure state with $\dim\mathcal H_B=d$.
Furthermore, let $\Pi$ denote the support projector of $\operatorname{Tr}_B\left(|\phi\rangle\langle\phi|\right)$, whose rank is at most $d$. Since
$|\phi\rangle\langle\phi|\leqslant \Pi\otimes\mathbb{I}_B$,
we have the following inequalities 
\begin{equation} \label{ineq}
\langle\phi|\rho_{AB}|\phi\rangle
\leqslant\operatorname{Tr}(\Pi\rho_A)
\leqslant\sum_{i=1}^{d}p_i=P_d.
\end{equation}
The last inequality follows from Ky Fan's maximum principle\footnote{See Theorem 6 of Ref.\cite{NielsenVidal}.}.

This bound can be attained as follows. First, transfer the $d$ most probable Schmidt components into orthogonal states of $B$, all accompanied by the same state of $D$ (the degrees of freedom outside $B$). The remaining components are distinguished by mutually orthogonal states of $D$, also orthogonal to that common state. The pure reference state obtained by keeping and renormalizing the first $d$ components is
\begin{equation}
 |\phi_d\rangle_{AB}=\frac{1}{\sqrt{P_d}}\sum_{i=1}^d\sqrt{p_i}\,|i\rangle_A|i\rangle_B.
\end{equation}
The actual $AB$ state may be mixed after $D$ is traced out, but its squared fidelity with this pure reference state is $\langle\phi_d|\rho_{AB}|\phi_d\rangle=P_d$.

For a state $\rho_{AB}$ and a pure state $|\phi\rangle_{AB}$, the squared fidelity is defined by $F^2\equiv\langle\phi|\rho_{AB}|\phi\rangle$. Inequality \eqref{ineq} therefore gives $F^2\leqslant P_d$. Since this bound is attainable by a suitable isometry, the final state of $AB$ can have squared fidelity at least $1-\varepsilon$ with some pure state if and only if
\begin{equation}
d\geqslant d_\varepsilon\equiv\min\left\{k~\middle|~\sum_{i=1}^{k}p_i\geqslant1-\varepsilon\right\}.
\label{oneShot}
\end{equation}
This spectral criterion is closely related to the construction involving a high-weight subspace used in quantum source coding\footnote{See Sec.V, Lemmas 1 and 2, and Sec.VII of Ref.\cite{Schumacher}.}. Unlike the
quantum source coding theorem \cite{Schumacher}, which concerns asymptotic compression of many identically prepared states, Eq.\eqref{oneShot} concerns a single bipartite state and gives a sharp dimension requirement for achieving a prescribed overlap with some pure target state. At leading order, the corresponding exact and finite-fidelity energy bounds follow by replacing $s$ in Eq.\eqref{energybound} by $\log r$ and $\log d_\varepsilon$, respectively. For a \emph{specified} target such as a thermofield double, compatibility of the reduced state, energy distribution and additional physical constraints are also required, i.e. the spectral criterion of Eq.(\ref{oneShot}) is not necessarily sufficient for the physical construction of a second black hole. In particular, we caution readers that equations \eqref{leakage} and \eqref{energybound}  are both entropy-level statements and should not be used in place of Eq.\eqref{oneShot} when exact or finite-error performance matters.

\section{Illustration using the Page curve}
After the Page time, the entanglement entropy $s\simeq S_{\rm BH}(M)=4\pi M^2$, so the capacity condition $s\leqslant S_{\rm BH}(m)$ requires $m\geqslant M$. Without external energy, the second black hole mass must be bounded by $m\leqslant E_R\simeq M_0-M$. Both conditions can hold only for $M\leqslant M_0/2$. After the Page time, but while $M>M_0/2$, some correlations with $A$ \emph{must} therefore remain outside $B$. For $M\leqslant M_0/2$, the entropy-capacity obstruction can be removed without external energy. However, this does not guarantee that a clean conversion is physically realizable.

For a large, initially pure Schwarzschild black hole emitting essentially photons and gravitons, the radiation entropy has been estimated by Page~\cite{Page}~(see Eq.(12) therein). 
Expressed in terms of the remaining mass fraction $x=M/M_0$, we get
\begin{equation}
\frac{s(x)}{4\pi M_0^2}\simeq\min\{\beta(1-x^2),x^2\},\quad x\equiv M/M_0,\quad\beta\simeq1.4847.
\label{Page}
\end{equation}
Here $\beta$ is the ratio of the semiclassical radiation entropy production rate to the magnitude of the black-hole entropy loss rate.
Inserting Eqs.\eqref{ER} and \eqref{Page} into the entropy-level energy requirement yields 
\begin{equation}
\frac{\mathcal{E}^{(S)}_{\rm req}}{M_0}\simeq\left[\sqrt{\min\{\beta(1-x^2),x^2\}}-(1-x)\right]_+.
\label{curve}
\end{equation}

The two branches of Eq.\eqref{Page} meet at the Page time, for which the normalized mass is
\begin{equation}
x_{\rm Page}=\sqrt{\frac{\beta}{1+\beta}} \simeq 0.773005.
\end{equation}
For $x\leqslant x_{\rm Page}$, one has $s\simeq4\pi M^2$, so that
\begin{equation}
\mathcal{E}^{(S)}_{\rm req}\simeq[2M-M_0]_+.
\label{postPage}
\end{equation}
Within this approximate model, Eq.\eqref{curve}
attains its maximum there, and
\begin{equation}
\left.\frac{\mathcal{E}^{(S)}_{\rm req}}{M_0}\right|_{\rm Page}
=2x_{\rm Page}-1\simeq0.546010.
\label{PageNum}
\end{equation}
By this point, the collected radiation carries only $E_R\simeq0.227M_0$, whereas the minimum entropy-level mass required for the second black hole is $m\simeq M\simeq0.773M_0$.

Alternatively, suppose that $\mathcal E_{\rm ext}=0$, so that $m\leqslant E_R\simeq M_0(1-x)$. Eq.\eqref{leakage} then gives the inequality
\begin{equation}
\frac{I(A\!:\!D)}{2s}\geqslant \left[1-\frac{4\pi M_0^2(1-x)^2}{s(x)}\right]_+.
\end{equation}
After the Page time $x\leqslant x_{\rm Page}$, the radiation entropy is approximated by that of the remaining black hole, $s(x)\simeq4\pi M_0^2x^2$. Hence, the previous inequality becomes
\begin{equation}
\frac{I(A\!:\!D)}{2s}\geqslant \left[1-\left(\frac{1-x}{x}\right)^2\right]_+.
\label{leakagePage}
\end{equation}
At $x=x_{\rm Page}$, the right-hand side is $0.913768$. Thus at least $91.4\%$ of the \emph{mutual-information budget} $2s$ must remain between the surviving hole and degrees of freedom outside $B$ under the stated capacity assumption. We caution readers that this does not quantify entanglement distillable from $D$. It also does not rule out a partially entangled $AB$ pair.

\section{Purification does not imply a geometric bridge}
Even if equations \eqref{rank}--\eqref{energybound} are satisfied, Alice has not necessarily prepared a thermofield-double (TFD) state: operations on $R$ alone leave $\rho_A$ unchanged, i.e.
$
\rho_A^{\mathrm{final}}=\rho_A^{\mathrm{initial}}.
$
A specified TFD state has the schematic form\footnote{An asymptotically flat Schwarzschild TFD requires a careful treatment, since the canonical ensemble is unstable; see, e.g., Ref.\cite{2309.05041} for a microcanonical construction.}
\begin{equation}\label{TFD}
|\mathrm{TFD}\rangle\propto\sum_n\ee^{-\beta_T E_n/2}|n\rangle_A|n\rangle_B.
\end{equation}
Such a state requires\footnote{The notations are standard: $H_A$ is the Hamiltonian of the subsystem $A$ such that $H_A|n\rangle_A=E_n|n\rangle_A$, and $Z=\operatorname{Tr}(e^{-\beta_T H_A})$ is the partition function with $\beta_T$ denoting the inverse temperature.} $\rho_A^{\mathrm{initial}}=\ee^{-\beta_T H_A}/Z$, unless additional operations on $A$ are allowed.

In their discussion of restoring the thermofield state, Maldacena and Susskind assume that the density matrix of each black hole is thermal~\cite{MaldacenaSusskind}. More generally, satisfying the entropy and Schmidt-rank conditions, even with substantial mutual information, does not guarantee the specific two-sided correlations required for a smooth ER-bridge geometry. A generic entangled state may still be compatible with ER=EPR without resembling the semiclassical smooth ER-bridge.

A further subtlety concerns the identification of the quantum subsystems themselves. For example, within a three-dimensional TQFT, ``fake entanglement'' can arise when apparently distinct degrees of freedom are identified~\cite{BaezVicary}. Such considerations lie beyond the scope of our work.

\section{Conclusion: Who Pays for \\the Bridge?}

The proposal to collect Hawking radiation from a black hole and convert it into a second black hole raises a natural question of thermodynamic bookkeeping: how are energy and quantum correlations redistributed during the process? Much like Maxwell's demon, an observer endowed with extraordinary control over a quantum system cannot dispense with the physical requirements of the operations performed. In the present thought experiment, possessing a purification of the original black hole is not equivalent to realizing that purification in a second gravitating system. 
The existence of a semiclassical Einstein--Rosen bridge, however, remains a separate question. Thus, the \emph{daemon sub ponte} does not challenge ER=EPR. Nevertheless, it reminds us that even an idealized preparation must respect the physical conditions of its realization. After all, even if ER=EPR holds, the energy and information-theoretic costs of constructing the second black hole must still be accounted for.\\
\begin{center}
\emph{The devil is in the details.}
\end{center}


\begin{thebibliography}{99}
\bibitem{MaldacenaSusskind} Juan Maldacena, Leonard Susskind, ``Cool Horizons for Entangled Black Holes'', {\hypersetup{urlcolor=vividviolet}\href{https://onlinelibrary.wiley.com/doi/10.1002/prop.201300020}{Fortschr. Phys. \textbf{61} (2013) 781}}, \href{https://arxiv.org/abs/1306.0533}{[arXiv:1306.0533 [hep-th]]}.


\bibitem{MaldacenaStanfordYang}
Juan Maldacena, Douglas Stanford, Zhenbin Yang, ``Diving Into Traversable Wormholes'', {\hypersetup{urlcolor=vividviolet}\href{https://onlinelibrary.wiley.com/doi/10.1002/prop.201700034}{Fortschr. Phys. \textbf{65} (2017) 5, 1700034}}, \href{https://arxiv.org/abs/1704.05333}{[arXiv:1704.05333 [hep-th]]}.

\bibitem{Boeretal}
Jan de Boer, Rik van Breukelen, Sagar F. Lokhande, Kyriakos Papadodimas, Erik Verlinde, ``Probing Typical Black Hole Microstates'', {\hypersetup{urlcolor=vividviolet}\href{https://link.springer.com/article/10.1007/JHEP01(2020)062}{JHEP \textbf{01} (2020) 062}}, \href{https://arxiv.org/abs/1901.08527}{[arXiv:1901.08527 [hep-th]]}.

\bibitem{Brownetal}
Adam R. Brown, Hrant Gharibyan, Geoff Penington, Leonard Susskind, ``The Python’s Lunch: Geometric Obstructions to Decoding Hawking Radiation'', {\hypersetup{urlcolor=vividviolet}\href{https://link.springer.com/article/10.1007/JHEP08(2020)121}{JHEP \textbf{08} (2020) 121}}, \href{https://arxiv.org/abs/1912.00228}{[arXiv:1912.00228 [hep-th]]}.


\bibitem{HarlowHayden} Daniel Harlow, Patrick Hayden, ``Quantum Computation vs. Firewalls'', {\hypersetup{urlcolor=vividviolet}\href{https://link.springer.com/article/10.1007/JHEP06(2013)085}{JHEP \textbf{06} (2013) 085}}, \href{https://arxiv.org/abs/1301.4504}{[arXiv:1301.4504 [hep-th]]}.

\bibitem{RemmenBaoPollack} Grant N. Remmen, Ning Bao, Jason Pollack, ``Entanglement Conservation, ER=EPR, and a New Classical Area Theorem for Wormholes'', {\hypersetup{urlcolor=vividviolet}\href{https://link.springer.com/article/10.1007/JHEP07(2016)048}{JHEP \textbf{07} (2016) 048}}, \href{https://arxiv.org/abs/1604.08217}{[arXiv:1604.08217 [hep-th]]}.


\bibitem{Bekenstein} Jacob D. Bekenstein, ``Black Holes and Entropy'', {\hypersetup{urlcolor=vividviolet}\href{https://journals.aps.org/prd/abstract/10.1103/PhysRevD.7.2333}{Phys. Rev. D \textbf{7} (1973) 2333}}.

\bibitem{Wall} Aron C. Wall, ``A Proof of the Generalized Second Law for Rapidly Changing Fields and Arbitrary Horizon Slices'', {\hypersetup{urlcolor=vividviolet}\href{https://journals.aps.org/prd/abstract/10.1103/PhysRevD.85.104049}{Phys. Rev. D \textbf{85} (2012) 104049}}; erratum {\hypersetup{urlcolor=vividviolet}\href{https://journals.aps.org/prd/abstract/10.1103/PhysRevD.87.069904}{\textbf{87} (2013) 069904}}, \href{https://arxiv.org/abs/1105.3445}{[arXiv:1105.3445 [gr-qc]]}.



\bibitem{NielsenVidal}
Michael Nielsen, Guifre Vidal,
``Majorization and the Interconversion of Bipartite States'',
{\hypersetup{urlcolor=vividviolet}\href{https://www.rintonpress.com/journals/doi/QIC1.1-5.html}{Quantum Inf. Comput. \textbf{1} (2001) 76}}.


\bibitem{Schumacher} Benjamin Schumacher, ``Quantum Coding'', {\hypersetup{urlcolor=vividviolet}\href{https://journals.aps.org/pra/abstract/10.1103/PhysRevA.51.2738}{Phys. Rev. A \textbf{51} (1995) 2738}}.

\bibitem{Page} Don N. Page, ``Time Dependence of Hawking Radiation Entropy'', {\hypersetup{urlcolor=vividviolet}\href{https://iopscience.iop.org/article/10.1088/1475-7516/2013/09/028}{JCAP \textbf{09} (2013) 028}}, \href{https://arxiv.org/abs/1301.4995}{[arXiv:1301.4995 [hep-th]]}.


\bibitem{2309.05041}
Wan Zhen Chua, Thomas Hartman, ``Black Hole Wavefunctions and Microcanonical States'', {\hypersetup{urlcolor=vividviolet}\href{https://link.springer.com/article/10.1007/JHEP06(2024)054}{JHEP \textbf{06} (2024) 054}}, \href{https://arxiv.org/abs/2309.05041}{[arXiv:2309.05041 [hep-th]]}.

\bibitem{BaezVicary}
John C. Baez, Jamie Vicary, ``Wormholes and Entanglement'', {\hypersetup{urlcolor=vividviolet}\href{https://iopscience.iop.org/article/10.1088/0264-9381/31/21/214007}{Class. Quant. Grav. \textbf{31} (2014) 21, 214007}}, \href{https://arxiv.org/abs/1401.3416}{[arXiv:1401.3416 [gr-qc]]}.

\end{thebibliography}
\end{document}